\documentclass[twocolumn,aps,prb,longbibliography,superscriptaddress]{revtex4-2}

\usepackage[pdftex]{graphics}
\usepackage{graphicx}
\usepackage{hyperref}
\usepackage{xcolor}
\usepackage{physics}
\usepackage{subfig}
\usepackage{bm}
\usepackage{caption}
\usepackage{amsmath} 
\usepackage{amssymb} 
\usepackage{mathtools,dsfont}

\DeclareMathOperator{\CC}{\mathcal{C}}
\DeclareMathOperator{\CCd}{\mathcal{C}^{-1}}

\DeclareMathOperator{\TT}{\mathcal{T}}
\DeclareMathOperator{\TTd}{\mathcal{T}^{-1}}
\DeclareMathOperator{\Tp}{\mathcal{T}_+}

\begin{document}
		
\title{Disorder and non-Hermiticity in Kitaev spin liquids with a Majorana Fermi surface}

\author{Lukas R\o dland}
\affiliation{Department of Physics, Stockholm University, AlbaNova University Center, SE-106 91 Stockholm, Sweden}

\author{Carlos Ortega-Taberner}
\affiliation{School of Physics, Trinity College Dublin, College Green, Dublin 2, Ireland}
 \affiliation{Trinity Quantum Alliance, Unit 16, Trinity Technology and Enterprise Centre, Pearse Street, Dublin 2, Ireland}

\author{Megha Agarwal}
\affiliation{Department of Physics, Stockholm University, AlbaNova University Center, SE-106 91 Stockholm, Sweden}

\author{Maria Hermanns}
\affiliation{Department of Physics, Stockholm University, AlbaNova University Center, SE-106 91 Stockholm, Sweden}

\date{\today}
\begin{abstract}
We study the effect of disorder on Z$_2$ quantum spin liquids with a Majorana Fermi line (respectively surface in three dimensions) and show that depending on the symmetries that are preserved \emph{on average} qualitatively different scenarios will occur. 
 In particular, we identify the relevant non-Hermitian symmetries for which disorder will effectively split the Fermi line into two exceptional lines, with $\Re(E)=0$ states filling the area in between. 
We demonstrate the different scenarios using both toy models as well as large-scale numerical simulations. 
\end{abstract}

\maketitle

\section{Introduction}
Ever since the original proposal by Anderson \cite{ANDERSON1973153}, quantum spin liquids have continued to fascinate theorists and experimentalists alike \cite{balents2010spin}. 
They show several unexpected properties, such as the lack of magnetic ordering even at zero temperatures, the presence of long-range entanglement \cite{Kitaev2006topological,Levin2006detecting} or the fractionalization of spins. 
Since they necessarily are strongly correlated phases, much of our current theoretical understanding of quantum spin liquids has been driven by simple model systems, such as the toric code \cite{kitaev2003fault} or Kitaev's honeycomb model \cite{kitaev2006anyons}. 
The latter has gained an increasing amount of attention lately as it may be realized in certain materials \cite{Jackeli2009Mott}, see also \cite{hermanns2018physics} and references therein. 

The Kitaev honeycomb model is exactly solvable in terms of Majorana fermions hopping in a static emergent Z$_2$ gauge field \cite{kitaev2006anyons}. 
It allows for a variety of different quantum spin liquid phases --- in the following referred to as Kitaev spin liquids (KSL) --- depending on the underlying lattice \cite{Yang2007mosaic,obrien2016classification}. 
At the same time it is sufficiently simple to allow for the computation of many, experimentally accessible properties, such as the dynamical structure factor \cite{knolle2014dynamics} or neutron scattering signatures \cite{banerjee2016proximate}. 
Here, we study the effect of hopping disorder on KSLs with a Majorana Fermi surface, both in a two-dimensional lattice \cite{Yang2007mosaic} and a three-dimensional one \cite{Hermanns2014quantum}. 
We also comment on the interesting properties of the effective Hamiltonian, obtained after disorder average, which links to the field of non-Hermitian physics.

The effects of disorder are analytically studied using Green's function formalism, a powerful tool for examining quantum states in disordered systems.
Disorder typically leads to a non-Hermitian self-energy of the system, and hence a broadening of the spectral function. This self-energy is then integrated into an effective non-Hermitian Hamiltonian \cite{Ashida_non_2020, bergholtz_exceptional_2021,okuma_non-hermitian_2023} at low energies, achieved through the Feshbach projection method \cite{rotter_non-hermitian_2009,Ashida_non_2020,meden_mathcalpt-symmetric_2023}. The imaginary part of this Hamiltonian's eigenvalues is directly related to the inverse lifetime of quasiparticles and thus to the broadening of the spectral function \cite{kozii_non-hermitian_2017}, reflecting the dynamic effects of disorder in the spectrum.

The existence of exceptional points where the Hamiltonian is non-diagonalizable is an inherent feature of non-Hermitian systems \cite{bergholtz_exceptional_2021}. A Dirac point in two dimensions can either be split into a pair of exceptional points or an exceptional ring under a non-Hermitian perturbation \cite{berry_physics_2004,szameit_mathcalpmathcalt-symmetry_2011,kawabata_classification_2019, cerjan_experimental_2019, zhen_spawning_2015}. 
 Between two paired exceptional points there is a Fermi arc \cite{kozii_non-hermitian_2017, carlstrom_exceptional_2018,zhou_observation_2018, yoshida_non-hermitian_2018,nagai_dmft_2020,cayao_exceptional_2023}, where the real part of the energy vanishes. 
Another intriguing aspect of non-Hermitian systems is the splitting of discrete symmetries into two different types, due to the difference between transposition and conjugation for non-Hermitian operators \cite{Bernard2002, kawabata_classification_2019}, giving rise to unique `transpose-type' symmetries that are inherently non-Hermitian. 
The two variations of time-reversal and particle-hole symmetries create different pathways to achieving chiral symmetry.

In Refs.~\cite{yang_exceptionalSL_2021,yang_exceptional_2022} the authors studied the effect of bond disorder on a Kitaev quantum spin liquid in the gapless phase on a honeycomb lattice, where they found that each gapless point splits into two exceptional points for generic (time-reversal preserving) disorder and opened into an exceptional ring for disorder that preserved time-reversal only on average. 
Inside the exceptional ring, the real part of the energy eigenvalue vanishes. 
This area of $Re[E]=0$ states is often called Fermi surface. 
However, since we need the (original) concept of Fermi surface for the three-dimensional model, we will instead call it a \emph{Fermi area} for the remainder of this manuscript. 
% They found that they have a Fermi area inside the exceptional ring, where the real part of the energy vanishes. 
The Fermi area of Ref.~\cite{yang_exceptional_2022} was very small and hard to observe. 
In the following, we show a systematic way to obtain this effect in a much more robust way. 

In this paper, our focus is on the gapless phase of the Kitaev spin liquid on the square octagon model in two dimensions and on the hyperoctagon model in three dimensions \cite{Hermanns2014quantum}.
The non-trivial implementation of time-reversal symmetry in the square-octagon (hyperoctagon) model leads ~\cite{obrien2016classification} to a Majorana Fermi line (surface), which is inherently stable.
Any non-Hermitian perturbation can only deform the Majorana Fermi line, not destroy it.
This is the generic, though slightly boring, scenario that prevails if all the symmetries are broken. 
Here, we show that there are two additional, symmetry-protected scenarios that occur.
For instance, introducing a time-reversal invariant hopping disorder (which partially breaks translation symmetry on average) leads to the Majorana Fermi line splitting into two exceptional lines.
This splitting creates a large Fermi area located between the two rings, which is protected by chiral symmetry. 
This Fermi area can be interpreted as the broadening of the spectral function, where the amount of broadening, e.g., the size of the Fermi area, is related to the lifetime of the single-particle states.
When chiral symmetry is broken, the exceptional lines gap out to pairs of exceptional points that are connected by Fermi arcs at $Re[E]=0$.  

\emph{Outline of this manuscript}
In Sec.~\ref{section:2d_model} we introduce a non-Hermitian toy model on the extended square octagon lattice that shows the effect in a transparent way, followed by the symmetry analysis in \ref{sec:Symmetry_analysis}.
We also discuss another toy model in Section \ref{sec:onsite_imaginary} as an aid for the disorder analysis in Section \ref{section:Disorder}. 
We end our manuscript with a short section on the generalization to three-dimensional models and some concluding remarks. Most of the technical details as well as a short introduction to the relevant KSLs can be found in the appendices.

\section{2D Model Hamiltonian and Symmetries}
\label{section:2d_model}

\subsection{Simple model}\label{sec:simple_model}
\begin{figure}[ht]
\centering
\includegraphics[width=\columnwidth]{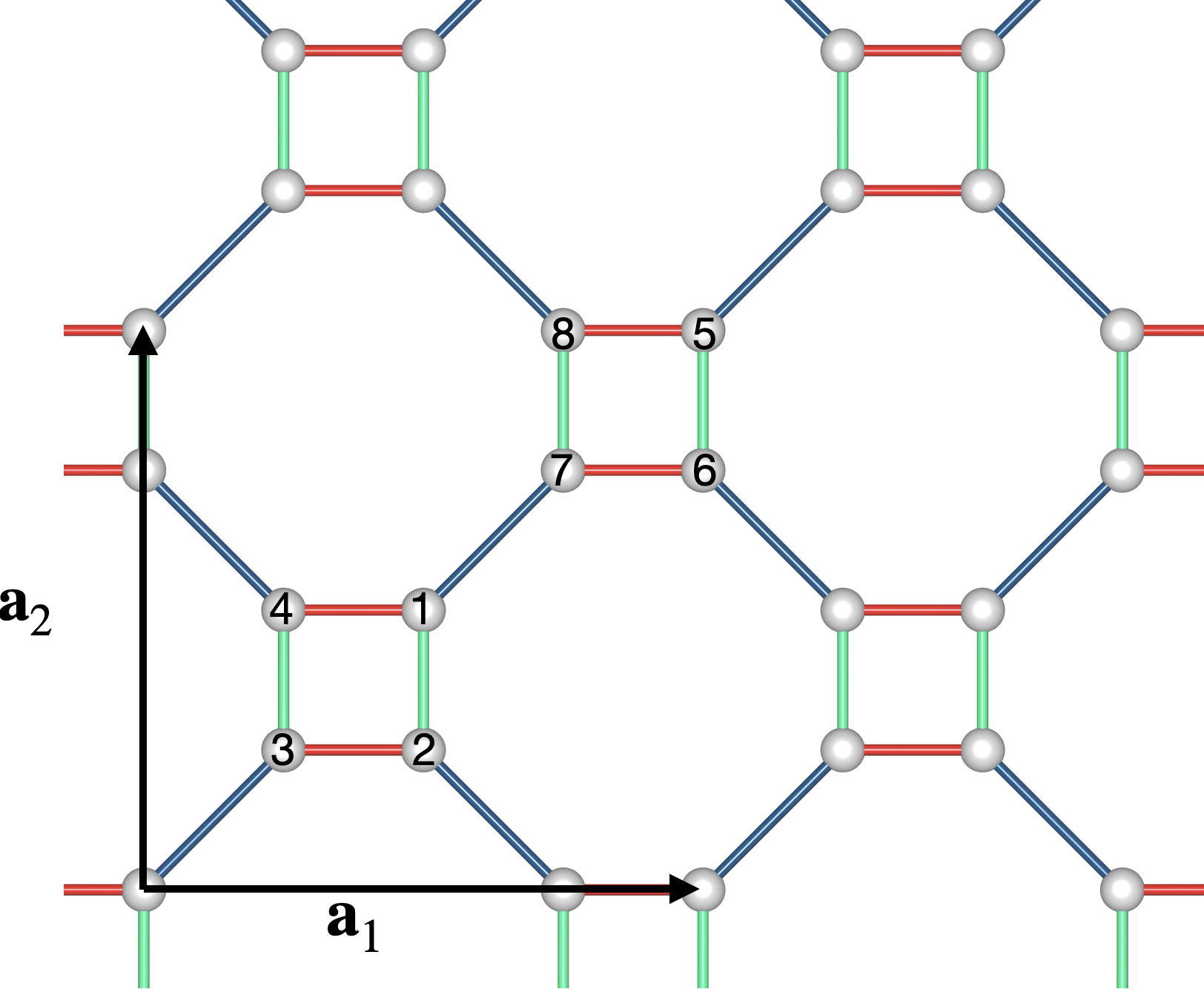}
\caption{Unit cell and translation vectors of the square-octagon model obtained after doubling the unit cell. Green/red/blue bonds correspond to nearest-neighbor Ising interactions with the Ising axis along $\hat x$/$\hat y$/$\hat z$. }
\label{fig:Model1Xband}
\end{figure}

We will consider a non-Hermitian version of the Kitaev spin liquid (KSL) on a square-octagon lattice. 
Readers unfamiliar with KSLs are referred to Appendix \ref{app:KSL}, which contains a short summary of KSLs in general as well as details for the two particular examples of KSLs relevant to this manuscript. 
The general solution of Kitaevs model \cite{kitaev2006anyons} relies on the fact that we can phrase the nearest neighbor Ising interaction of spin-$1/2$ local moments in terms of a nearest neighbor hopping model of Majorana fermions as $\mathcal{H}=i\sum J_\gamma u_{jk} c_jc_k$.
This model is Hermitian as long as $J_\gamma\in\mathbb R$. 
The square-octagon lattice as well as the choice of Ising interactions are depicted in Fig.~\ref{fig:Model1Xband}. The original square-octagon lattice has a 4-site unit cell, given by sites 1 to 4. 
In the following, we will partially break translation symmetry which doubles the unit cell in the way shown in Fig.~\ref{fig:Model1Xband}.

\begin{figure}[ht]
	\centering
	\includegraphics[width=\columnwidth]{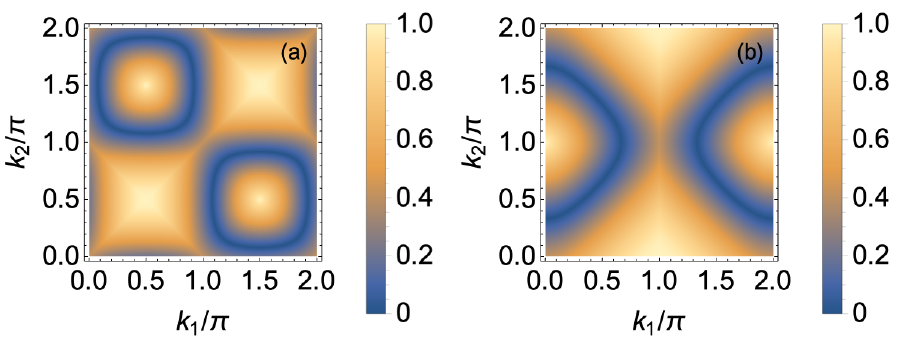}
	\caption{ Real part of the energy gap for the square octagon model with $\phi = \theta = 0 $ using the (a) 4 site unit cell and (b) the extended 8 site unit cell, where the two Majorana Fermi lines are mapped into each other.}
\label{fig:fig1}
\end{figure}

Using the 4-site unit cell, the Hermitian version of this model has two Fermi lines in the spectrum, shown in Fig.~\ref{fig:fig1}(a). 
A symmetry analysis along the lines of Ref.~\cite{obrien2016classification} shows that there is perfect nesting between the two surfaces, as long as time-reversal symmetry or inversion symmetry remain intact. 
In particular, they can be mapped onto each other by a translation of $\mathbf k_0$, where $\mathbf k_0$ is half a reciprocal lattice vector; see Appendix \ref{app:KSL}. 
Any translationally invariant two-Majorana term, even a generic non-Hermitian one, can only deform the Fermi lines.
In the absence of any symmetry, the Kitaev model on the square-octagon lattice will exhibit Majorana Fermi lines independent of the non-Hermitian perturbation we add. 
However, when retaining some symmetries, the scenario can become widely different. 
To see this, we need to enlarge the unit cell to contain 8 sites, as shown in Fig.~\ref{fig:Model1Xband}.
By doubling the unit cell, the Majorana Fermi lines are mapped onto each other, as depicted in Fig.~\ref{fig:fig1}(b). Furthermore, they are protected by the full translation symmetry, i.e., under translations $(\mathbf a_1 \pm \mathbf a_2)/2$.

We now consider perturbations that mimic a dimerization in the system, i.e., that break the original translation symmetry, but keep the translation symmetry of the 8-site model. 
These may lead to two different scenarios. 
Hermitian perturbations that preserve time-reversal but break inversion reduce the doubly degenerate Fermi line to an even number of Dirac points.  
When adding a non-Hermitian perturbation, the Dirac points will split into a pair of exceptional points --- a behavior already seen in Ref.~\cite{yang_exceptionalSL_2021}.
However, we can also imagine a much more interesting scenario where the Fermi line splits into two exceptional lines that delimit a Fermi area with $Re[E]=0$. 
Before discussing the symmetry that protects the exceptional rings, we first introduce a simple explicit model that showcases both scenarios. 

We introduce non-Hermiticity in the square-octagon model by allowing the couplings to acquire an additional phase, similar to what was considered in Ref.~\cite{yang_exceptionalSL_2021}, except that we allow the phases to be different in the two squares of the extended unit cell. 
In particular, let us consider the Hamiltonian given by 
\onecolumngrid
 \begin{align}
H =\left(
\begin{array}{cccccccc}
 0 & i J_x e^{i \phi } & 0 & i J_y & 0 & 0 & i J_z & 0 \\
 -i J_x e^{i \phi } & 0 & i J_y & 0 & 0 & 0 & 0 & i e^{-i k_2} J_z \\
 0 & -i J_y & 0 & -i J_x & -i e^{i \left(-k_1-k_2\right)} J_z & 0 & 0 & 0 \\
 -i J_y & 0 & i J_x & 0 & 0 & -i e^{-i k_1} J_z & 0 & 0 \\
 0 & 0 & i e^{i \left(k_1+k_2\right)} J_z & 0 & 0 & i J_x e^{i \theta} & 0 & i J_y \\
 0 & 0 & 0 & i e^{i k_1} J_z & -i J_x e^{i \theta} & 0 & i J_y & 0 \\
 -i J_z & 0 & 0 & 0 & 0 & -i J_y & 0 & -i J_x \\
 0 & -i e^{i k_2} J_z & 0 & 0 & -i J_y & 0 & i J_x & 0 \\
\end{array}
\right),
\label{eq:hamiltonian}
\end{align}
\twocolumngrid
\noindent 
where a phase $\phi$ was added in the hopping between sites $1$ and $2$, and a phase $\theta$ in the hopping between sites $5$ and $6$.
Note that the phase acquired is the same for the hopping in both directions, leading to non-Hermiticity. 

\begin{figure}[t]
	\centering
	\includegraphics[width=\columnwidth]{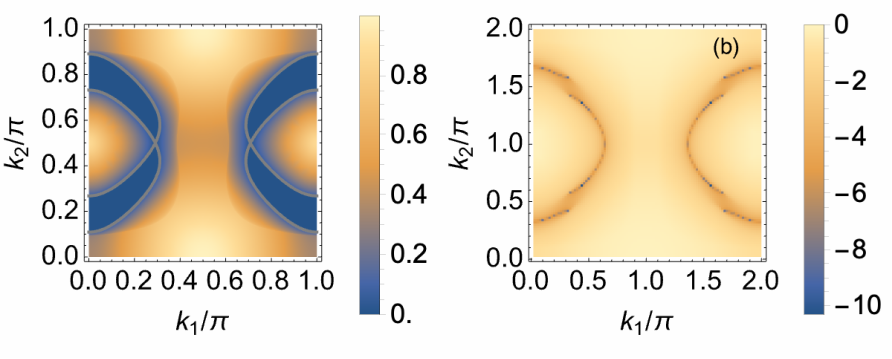}
	\caption{ (a) Real part of the energy gap for the non-Hermitian model using $\phi = 1/2 = -\theta$, showing Fermi areas. Exceptional lines are indicated by the gray boundaries. (b) The logarithm of the real part of the gap for the model with $\phi = 1/2,\theta = 0$, where the exponential rings split into points at the ends of Fermi arcs with $Re[E]=0$.}
\label{fig:fig2}
\end{figure}

If the two phases are exactly opposite, i.e., $\phi=-\theta$, the resulting non-Hermitian perturbation indeed splits the Majorana Fermi lines into two exceptional rings, enclosing a Fermi area of zero real energy states, as can be seen in Fig.~\ref{fig:fig2}(a). 
Moving away from $\theta = -\phi$, as shown in Fig.~\ref{fig:fig2}(b), causes the exceptional rings to gap into isolated exceptional points that delimit Fermi arcs.
In the following section, we will show that the existence of exceptional rings is not fine-tuned, as one might suspect from our model system, but symmetry-protected.
Even the second scenario, where we find isolated exceptional points at zero energy, is protected by a symmetry, albeit a weaker one. 
While the exceptional rings require time-reversal symmetry (alternatively inversion), the latter relies on sublattice symmetry (with broken time-reversal and inversion), see Appendix \ref{app:symmetries}.
Note that time-reversal and sublattice symmetry cannot be treated as two independent symmetries for Hermitian Kitaev spin liquids; for non-Hermitian models, as well as for the effective Hamiltonian of the disordered systems, they are distinct.

\subsection{Symmetry analysis}\label{sec:Symmetry_analysis}
%The exceptional lines break up into points when $\theta\neq -\phi$, which implies that the Fermi surface disappears. 
We will now proceed to show that a version of chiral symmetry is responsible for the protection of the exceptional lines. 
Its action on the Hamiltonian is given by
\begin{equation}
    U H^\dagger(\mathbf k) U^{\dagger} = -H(\mathbf k),
    \label{eq:chiral_symmetry1}
\end{equation}
where $U$ is unitary. 
This means that the Hamiltonian is anti-pseudo-hermitian, that is, equal to $iH_{PsH}$, for some pseudo-Hermitian $ H_{PsH}$. Pseudo-Hermitian Hamiltonians possess eigenvalues that are either completely real or form complex conjugate pairs \cite{mostafazadeh_pseudo-hermiticity_2002}. Thus, the eigenvalues of an anti-pseudo-Hermitian Hamiltonian can be either purely imaginary or occur in pairs with opposite real parts. This difference causes the system to divide into two regions. The first region has zero real energy parts, while the second displays opposite real energy parts. Notably, these regions are separated by exceptional lines \cite{mostafazadeh_pseudo-hermiticity_2002}. The anti-pseudo-Hermitian symmetry, therefore, protects the Fermi areas where the real part of the energy vanishes.

Chiral symmetry is typically achieved through a combination of particle-hole symmetry and time-reversal symmetry \cite{PhysRevX.9.041015}.  In our system, particle-hole symmetry is inherently present due to the Majorana representation, as described by the equation:
\begin{align}\label{eq:chiral_sym}
        \CC_{-} H^T(\mathbf k) \CCd_{-}=-H(-\mathbf k),
\end{align}
with $\CC_{-} = \text{diag}\left(
1,1,1,1,1,1,1,1
\right)$. 

Conversely, the implementation of time-reversal symmetry, though standard, is less straightforward compared to particle-hole symmetry. In the many-body framework, time-reversal symmetry is defined by:
\begin{align}
    \Tilde{\TT}_+ \mathcal{H}\Tilde{\TT}_+^{-1}=\mathcal{H},
\end{align}
where $\mathcal{H}$ is the second quantized Hamiltonian and $\Tilde{\TT}_+$ is an anti-unitary operator. This operator acts on the creation operators $c_j$ as follows:
\begin{align}
    \Tilde{\TT}_+c_j\Tilde{\TT}_+^{-1}= \begin{cases}
    c_j,\quad j\in A  \nonumber\\
    -c_j,\quad j\in B.
    \end{cases}
\end{align}
Applying this to the many-body Hamiltonian,  $\mathcal{H}=i\sum J_\gamma u_{jk} c_jc_k$ yields:
\begin{align}
    \Tilde{\TT}_+ \mathcal{H}\Tilde{\TT}_+^{-1}=i\sum J_\gamma^* u_{jk} c_jc_k\neq \mathcal{H}  \text{  for } J_\gamma\neq J_\gamma^*,
\end{align}
and therefore any non-Hermiticity we consider will break this symmetry. However, in our model, the symmetry can be restored by setting $\theta = -\phi$ and supplementing time reversal with a half-unit cell translation:
\begin{align}\label{eq:tr_with_phase}
    (\Tp U_{1/2})H^*(-\mathbf k) (\Tp U_{1/2})^{-1}=H(\mathbf k),
\end{align}
where 
 \begin{align}\label{eq:Tplus}
    \Tp=\text{diag}\left(
1,-1,1,-1,-1,1,-1,1
\right)
\end{align}
 and
\begin{align}\label{eq:U1/2}
   U_{1/2}= \left(
\begin{array}{cc}
 0 & \mathds{1}_{4\times 4} \\
 e^{i (  k_1+k_2)}\mathds{1}_{4\times 4} & 0  \\
\end{array}
\right).
\end{align}

Combining Eq.~\eqref{eq:tr_with_phase} with particle-hole symmetry from Eq.~\eqref{eq:chiral_sym} we obtain the symmetry~\eqref{eq:chiral_symmetry1} with $U =  \CC_{-}\Tp U_{1/2} $ which protects the Fermi area.
The particle-hole symmetry is inherent in the Majorana formulation, so the protection of the Fermi area hinges on the combined half-translation and time-reversal symmetry.

There are various variations of Hamiltonians \eqref{eq:hamiltonian} that keep above symmetry and, thus, exhibit exceptional lines. Obviously, we could have added phases with opposite signs for other bonds, e.g., between sites 2 and 3, and between 6 and 7. 
Another alternative is to make the hopping strength non-reciprocal, e.g., by changing the hopping from sites 1 and 7 to $J_z (1+\delta_1)$, the one from sites 3 and 5 to $J_z (1+\delta_1^*)$, while keeping the hopping from 7 to 1, as well as from 5 to 3, unchanged. 
Not all of the possibilities to keep the symmetry \eqref{eq:chiral_symmetry1} are physical in the system at hand, but they may be relevant for other realizations.
In the next subsection, we discuss another way to make the Hamiltonian non-Hermitian while retaining \eqref{eq:chiral_symmetry1} that at first glance seems unphysical, but is closely related to what happens when considering disorder in the model.

We can illustrate Fermi areas by observing how chiral symmetry maps states within these areas to themselves, while states outside are mapped to orthogonal states. This results in an overlap quantized to either 1 or 0. The formula for calculating this overlap is:
\begin{align}
\bra{\psi^R}U^{-1}\ket{\psi^R} \frac{\norm{\psi^L}}{\norm{U^{-1}\ket{\psi^R}}},
\label{eq:sym_state}
\end{align}
as outlined in Appendix \ref{Section:Anti-ph}. This technique effectively differentiates between regions inside and outside the Fermi area, offering a clear visualization of the $Re[E]=0$ surfaces.  This is illustrated in Fig.~\ref{fig:fig4} (b), for adding the onsite imaginary terms discussed in the following section. Additionally, Fig.\ref{fig:fig3} in Appendix \ref{Section:Anti-ph} showcases the self-symmetric states for the model Hamiltonian \eqref{eq:hamiltonian} when $\phi=-\theta =1/2$.

\subsection{On-site imaginary terms}\label{sec:onsite_imaginary}

In the following section, we study how non-Hermiticity appears in an effective description when disorder is introduced. However, non-Hermitian perturbations like the ones in Eq.~\eqref{eq:hamiltonian} are difficult to obtain using this procedure. Typically, non-Hermiticity in this context appears as imaginary on-site potentials. Before addressing the full disorder calculation in the next section, we will model this effect by setting $\phi=\theta =0$ in Eq.~\eqref{eq:hamiltonian} and adding a diagonal term 
\begin{equation}
    \Sigma = i \textrm{diag}(s_1,s_2,...,s_8),
\end{equation}
where $s_j \in \mathbb{R}$. Note that diagonal terms such as this do not make sense from the Majorana perspective, as $i s c_j c_j = 0$ \cite{footnote1}. However, as we will see later, they appear in the Green's function description and have relevant consequences.

Both particle-hole symmetry and time-reversal symmetry split into two distinct symmetry classes for non-Hermitian systems \cite{PhysRevX.9.041015}, due to a distinction between transposition and complex conjugation for non-Hermitian operators. Adding imaginary terms to the diagonal breaks both time-reversal and particle-hole symmetry as defined in Sec.~\ref{sec:Symmetry_analysis}, but it does not break the transpose type symmetries TRS$^\dagger$, given by
\begin{align}\label{eq:TRS_transpose}
        &\CC_{+} H^T(\mathbf k) \CCd_{+}=H(-\mathbf k)  \nonumber\\
        &\CC_{+} = \text{diag}\left(1,-1,1,-1,-1,1,-1,1\right),
\end{align}
or PHS$^\dagger$, given by
\begin{align}
        &\TT_{-} H^*(\mathbf k) \TTd_{-}=-H(-\mathbf k) \nonumber\\
        &\TT_{-} = \text{diag}\left(1,1,1,1,1,1,1,1\right).
\end{align}
The combination of both symmetries results again in a chiral symmetry required to protect Fermi areas,
\begin{align}
    &U H^\dagger(\mathbf k) U^{-1} = -H(\mathbf k),  \nonumber\\
    &U = \CC_+ \TT_- .
    \label{eq:chiral_symmetry_2}
\end{align}

\begin{figure}[t]
	\centering
	\includegraphics[width=\columnwidth]{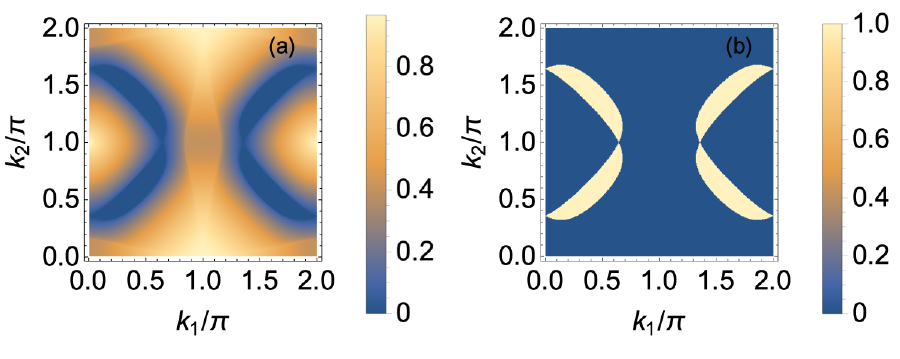}
	\caption{ (a) Real part of the energy gap for the non-Hermitian model with on-site imaginary terms on one square with $s = 1/2$, showing Fermi areas. (b) The corresponding self-symmetric states from Eq.~\eqref{eq:sym_state}.}
\label{fig:fig4}
\end{figure}

To split the Fermi line into exceptional lines, we need to, as discussed in Sec.~\ref{sec:simple_model}, either break time-reversal symmetry or the translation of the original 4-site unit cell model. Since adding imaginary terms to the diagonal preserves a form of time-reversal symmetry, it becomes necessary to break the half-translation symmetry $U_{1/2}$. The half-translation symmetry is broken when the onsite term is of different strength on different squares, that is, $s_j\neq s_{j+4}$. The extent of the Fermi area depends on the degree to which the half-translation symmetry is broken. By choosing $\Sigma =i s \times \text{diag}\left(1,1,1,1,0,0,0,0\right)$, the effect is maximized, but the results remain qualitatively the same regardless of the specific choice of $\Sigma$, as long as the original translation symmetry $U_{1/2}$ is broken.
The spectrum resulting from this non-Hermitian term is shown in Fig.~\ref{fig:fig4}(a), using $s = 1/2$.
The system hosts Fermi areas with $Re[E]=0$ that are protected by the symmetry in Eq.~\eqref{eq:chiral_symmetry_2}, as illustrated in Fig.~\ref{fig:fig4}(b).

\section{Disorder}\label{section:Disorder}
In this section, we investigate the impact of introducing disorder to the nearest-neighbor bond hoppings in the Hermitian KSL on a square-octagon lattice. Our analysis is confined to disordering the hoppings along bonds that lie within the 8-site unit cell. 
While this assumption is not essential for the effect, it makes the analysis simpler.
Including only nearest neighbor hopping disorder ensures that time-reversal symmetry, in the form of Eq.\eqref{eq:TRS_transpose}, remains intact, which is crucial for protecting the Fermi area.

The Hamiltonian for this system can be expressed as:
\begin{align}
    H =& H_0 + V \nonumber \\
    =& \sum_{\boldsymbol{r}\boldsymbol{r}',\alpha \beta.}t_{\alpha\beta} (\boldsymbol{r}-\boldsymbol{r}')\gamma_{\alpha}(\boldsymbol{r}) \gamma_{\beta}(\boldsymbol{r}')\nonumber \\ 
    + &\sum_{\boldsymbol{r},\alpha\beta} V_{\alpha\beta}(\boldsymbol{r})\gamma_{\alpha}(\boldsymbol{r}) \gamma_{\beta}(\boldsymbol{r}),
\end{align}
where $\boldsymbol{r} $ denotes the unit cell, and $\alpha, \beta,\dots$ represents the position within the unit cell. 
The term \( t_{\alpha\beta} (\boldsymbol{r} - \boldsymbol{r}') \) denotes the hopping strength between sites \( \alpha \) and \( \beta \) in the Hamiltonian. 
The disorder term \( V_{\alpha\beta}(\boldsymbol{r}) \) characterizes the randomness in the hopping along the bond between sites \( \alpha \) and \( \beta \). 
The hopping, and thus also the disorder $V$, is only nonzero for nearest neighbors in our model. 
The average value of this disorder term is assumed to be zero, i.e., \( \langle V_{\alpha \beta} \rangle = 0 \). 
This assumption is based on the premise that any non-zero average disorder could be effectively incorporated into the hopping parameters \( t_{\alpha \beta} \), thus altering the baseline hopping strengths in the Hamiltonian. 
Note that equal hopping strength $J_x = J_y= J_z$ is not at all important for our analysis.

The KSL with disorder can be effectively described by a non-Hermitian Hamiltonian using the Feshbach projection \cite{rotter_non-hermitian_2009, yang_exceptional_2022}.
The Green's function in the momentum space of a disordered system is given by 
\begin{align}
    G(\mathbf{k},\omega) = \frac{1}{\omega - (H_0(\mathbf{k}) + \Sigma(\mathbf{k},\omega))}.
\end{align}
Here $\Sigma (\mathbf{k},\omega)$ is the self-energy of the system, which captures the information from the disorder. The effective Hamiltonian is given by $H_{eff}(\mathbf{k}, \omega)=H_0(\mathbf{k})+\Sigma(\mathbf{k}, \omega)$, which at low energies can be approximated as $H_{eff}(\mathbf{k}):=H_{eff}(\mathbf{k},0)$, which in general is a non-Hermitian operator. 

The self-energy is formally given by
\begin{align}
    \Sigma(\mathbf{k},\omega)= g^{-1}(\mathbf{k}, \omega)-G^{-1}(\mathbf{k}, \omega),
\end{align}
where $g$ is the retarded Green's function for the system without disorder. Employing the self-consistent Born approximation \cite{yang_exceptional_2022} reveals the self-energy resulting from this disorder. Importantly, this approximation gives an iterative method to calculate the self-energy. In Appendix \ref{section:SCBA} we show that the self-energy for disorder within one unit cell can be expressed as:
\begin{align}\label{eq:SelfEnergyDef}
    \Sigma_{\alpha \beta}(\bold k, \omega)= \sum_{\lambda\delta} F_{\alpha\lambda\delta\beta} G_{\lambda \delta}(\bold k, \omega),
\end{align}
where $F_{\alpha\lambda\delta\beta}=\left< V_{\alpha \lambda} V_{\delta\beta} \right>$ captures the characteristics of the disorder.

\begin{figure}[t]
	\centering
	\includegraphics[width=\columnwidth]{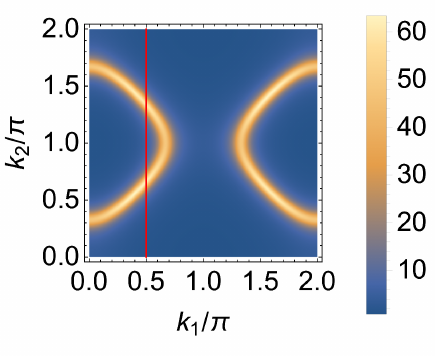}
\caption{Spectral function at zero frequency, showing the annulus Fermi area. The red line shows the cut in momentum space used in Fig.~\ref{fig:spec1234disorder}.}
\label{fig:spec1234disorder2}
\end{figure}

\begin{figure}[t]
	\centering
	\includegraphics[width=\columnwidth]{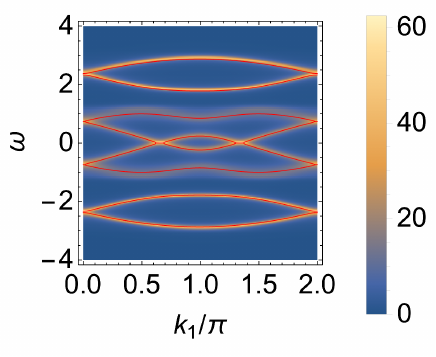}
	\caption{Spectral function of the disordered system at $k_1  = \frac\pi 2$ and the real part of energy bands of the effective Hamiltonian in red. The Fermi area can be seen as the flat regions at zero frequency.}
\label{fig:spec1234disorder}
\end{figure}

We now describe one way to obtain Fermi areas in the effective Hamiltonian in a similar way as in the toy model with on-site imaginary terms. 
By adding random disorder on the x- and y-bonds, 
one gets a non-Hermitian contribution to the effective Hamiltonian only on the diagonal. 
One can break the half-translation symmetry by having a different disorder strength on the two squares of the unit cell, which breaks the Fermi line into exceptional lines. 
The Fermi area is,  therefore, protected by a symmetry of the form of equation \eqref{eq:chiral_symmetry_2}, and the area of the surface is related to the difference in disorder strength on the two squares, e.g., how strong the breaking of translation symmetry is. 
The Fermi area forms an annulus enclosed by exceptional lines.

The spectral function, defined as $A(\bold k,\omega) = -2\Im \Tr(G(\bold k,\omega))$, shows the density of states \cite{kozii_non-hermitian_2017}. 
In Fig.~\ref{fig:spec1234disorder2} we show the spectral function at $\omega = 0$, which shows the Fermi area as an annulus in momentum space with a high density of states. 
The red line in Fig.~\ref{fig:spec1234disorder2} indicates the cut in momentum space at $k_1=\frac\pi 2$ that is used in Fig.~\ref{fig:spec1234disorder} to show the spectral function at different frequencies and the real part of the eigenvalues of $H_{eff}$ (red line). 
Here one can see the Fermi area as a flat region at zero frequency.

\section{3D Model Hamiltonian}\label{sec:3Dmodel}
Up to this point, we solely discussed two-dimensional systems. 
But the same physics plays out in a similar fashion for three-dimensional ones. 
Here, the starting point is a system with a doubly-degenerate Majorana Fermi surface.
These can be realized in a variety of Kitaev spin liquids \cite{obrien2016classification}, but also in metals \cite{Bzdusek2017robust} or superconductors \cite{agterberg2017bogoliubov} in the presence of inversion symmetry. 
Making the system non-Hermitian while retaining the symmetry \eqref{eq:chiral_symmetry1} splits the Fermi surfaces into two exceptional surfaces that now delimit a volume of $\Re[E]=0$ states. 
A suitable model system to illustrate this is the Kitaev spin liquid on the hyperoctagon lattice \cite{Hermanns2014quantum}.  
The following discussion is very similar to the two-dimensional case.
Its main objective is to showcase an explicit, three-dimensional example.

The KSL on the hyperoctagon lattice harbors two Fermi surfaces that are again perfectly nested.
The Fermi surfaces are again stable against any translationally invariant two-Majorana term (with respect to the standard 4-site unit cell). 
The perfect nesting is protected by time-reversal symmetry. 
We now proceed as before, we double the unit cell and introduce a non-Hermitian perturbation that breaks the original translation symmetry in such a way that the symmetry \eqref{eq:chiral_symmetry1} remains intact.

For the sake of an explicit example, we consider imaginary onsite terms, as in Sec.~\ref{sec:onsite_imaginary}, for which the corresponding Bloch Hamiltonian is given by
\onecolumngrid
\begin{equation}
\left(
\begin{array}{cccccccc}
	i s & -i J_y & 0 & -i J_z e^{-i k_1} & 0 & 0 & -i J_x e^{-i (k_1+k_2)} & 0 \\
	i J_y & 0 & i J_z & 0 & 0 & 0 & 0 & -i J_x e^{-i (k_1+k_2+k_3)} \\
	0 & -i J_z & i s & i J_y & i J_x e^{-i k_3} & 0 & 0 & 0 \\
	i J_z e^{i k_1} & 0 & -i J_y & 0 & 0 & i J_x & 0 & 0 \\
	0 & 0 & -i J_x e^{i k_3} & 0 & 0 & -i J_y & 0 & -i J_z e^{-i k_1} \\
	0 & 0 & 0 & -i J_x & i J_y & i s & i J_z & 0 \\
	i J_x e^{i (k_1+k_2)} & 0 & 0 & 0 & 0 & -i J_z & 0 & i J_y \\
	0 & i J_x e^{i (k_1+k_2+k_3)} & 0 & 0 & i J_z e^{i k_1} & 0 & -i J_y & is , \\
\end{array}
\right)
\label{eq:hyperoctagon_h}
\end{equation}
\twocolumngrid
\noindent 
where the diagonal terms correspond to the non-Hermitian perturbation, with strength $s$.
Such a scenario is again, as seen in Sec.~\ref{section:Disorder}, related to the effective Hamiltonian of a disordered system. 
However, since disorder calculations are prohibitively expensive for three-dimensional systems, we restrict the discussion to these model systems. 

\begin{figure}[t]
	\centering
	\includegraphics[width=\columnwidth]{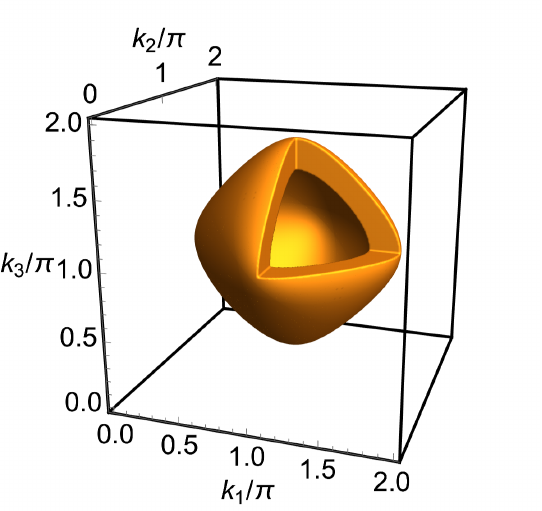}
\caption{Volume of states at $\Re[E]=0$ for the model in Eq.~\eqref{eq:hyperoctagon_h} with $s=1/4$.  }
\label{fig:hyperoctagon}
\end{figure}

 This Hamiltonian has the same non-spatial symmetries as the two-dimensional model with on-site imaginary terms studied above, TRS$^\dagger$
\begin{align}
        &\CC_{+} H^T(\mathbf k) \CCd_{+}=H(-\mathbf k)  \nonumber\\
        &\CC_{+} = \text{diag}\left(1,-1,1,-1,-1,1,-1,1\right),  \nonumber\\
        &\{ \epsilon(\mathbf k)\}=\{ \epsilon(-\mathbf k)\},
\end{align}
and PHS$^\dagger$,
\begin{align}
        &\TT_{-} H^*(\mathbf k) \TTd_{-}=-H(-\mathbf k) \nonumber\\
        &\TT_{-} = \text{diag}\left(1,1,1,1,1,1,1,1\right), \nonumber\\
        &\{ \epsilon^*(\mathbf k)\}=\{ -\epsilon(-\mathbf k)\}.
\end{align}
Combining both of these symmetries results again in a chiral symmetry that can protect volumes with $Re(E) = 0$,
\begin{align}
    &U H^\dagger(\mathbf k) U^{-1} = -H(\mathbf k),  \nonumber\\
    &U = \CC_+ \TT_- .
    \label{eq:chiral_symmetry_3}
\end{align}
In Fig.~\ref{fig:hyperoctagon} we show that this is the case, for the model in Eq.~\eqref{eq:hyperoctagon_h} with $s=1/4$. A cut has been made in the plot to better visualize that the region is indeed a volume in $\mathbf k$ space.

Note that, as in the square-octagon case, we have various ways to build non-Hermitian systems with such exceptional surfaces. 
We could instead have introduced phases on the hoppings or introduced nonreciprocal hoppings, as in Sec.~\ref{sec:simple_model}. 
In that case, the chiral symmetry requires an explicit half-translation $U_{1/2}$, which is obtained from Eq.~\eqref{eq:U1/2} by replacing $e^{i (k_1+k_2)} \rightarrow e^{i(k_1+k_2+k_3)}$. 
The symmetry in \eqref{eq:chiral_symmetry1} is fulfilled with $U =  \CC_{-}\Tp U_{1/2} $, with $\CC_{-}$ and $\Tp$ as given in \eqref{eq:chiral_sym} and \eqref{eq:Tplus}, respectively. 

\section{Conclusions}
We examined the effects of (weak) disorder on gapless Kitaev spin liquids in the square-octagon lattice in two dimensions and the hyperoctagon lattice in three dimensions, employing an effective non-Hermitian Hamiltonian derived from Green's function formalism.  Unlike the work done in \cite{yang_exceptional_2022}, where they considered a lattice with a Dirac point, we studied lattices where the codimension of gapless points is 1, that is, lines in two dimensions and surfaces in three dimensions.

The effect of disorder crucially depends on the kinds of symmetries that are either retained or broken on average.
Disorder that does not break translation symmetry on average will, of course, result in a Majorana metal, reflecting the stability of the Majorana Fermi surface.
Interesting effects appear when we consider disorder that induces `dimerization' in the system and treats the two squares in the 8-site unit cell differently.
Chiral symmetry protects pairs of exceptional lines that are connected by a Fermi area at $Re[E]=0$.
In this manuscript, we discuss the case where chiral symmetry is achieved by combining time-reversal and particle-hole symmetry.
However, for the square-octagon model, we could have equally well replaced time-reversal with inversion.
Broken chiral symmetry results in isolated exceptional points if sublattice symmetry is retained, or in Fermi lines if it is not.
The latter scenario can be most easily understood by noting that sublattice symmetry pins the exceptional points to $Re[E]=0$.
When it is also broken, the exceptional points move up or down in $Re[E]$, and the Fermi arcs of Fig.~\ref{fig:fig2} evolve into Fermi lines. 

There has been an extensive amount of work has been done on studying the topological properties of non-Hermitian systems in the last couple of years. One such topological phenomenon is the non-Hermitian skin effect~\cite{lee_anomalous_2016, kunst_biorthogonal_2018, martinez_alvarez_non-hermitian_2018, xiong_why_2018, yao_edge_2018}, where the bulk states are localized at the boundary of the system. The existence of isolated exceptional points directly leads to the skin effect~\cite{okuma_non-hermitian_2023}. Disorder that breaks time-reversal symmetry will therefore lead to an accumulation of modes on the boundary. Disorder that preserves time-reversal symmetry could therefore be clearly distinguished from symmetry-breaking disorder by studying the boundary of the system. Future work into effective non-Hermitian Hamiltonians could therefore give new insights into the topological phenomena of quantum spin liquids.

\emph{Acknowledgements }---  The research in this grant was supported by the Swedish Research Council under grant no. 2017-05162 and the Knut and Alice Wallenberg Foundation under grant no. 2017.0157.

\appendix

\section{Self-consistent Born approximation}\label{section:SCBA}

Here we derive the expression in equation \eqref{eq:SelfEnergyDef}, which is the self-energy for a system with disorder. 
For the sake of simplicity, we assume that the disorder only acts on bonds within the unit cell, not on those coupling neighboring unit cells. 
The retarded Green's function is defined as
\begin{align}
    G^r_{\alpha \beta}(t,t';\boldsymbol{r},\boldsymbol{r}') =-i\expval{\{ \gamma_{\alpha} (\boldsymbol{r},t), \gamma_{\beta} (\boldsymbol{r}',t')\} } \theta(t-t').
\end{align}
Before proceeding, it is convenient to compute the time evolution of the Majorana operators
\begin{align}
i\partial_t     \gamma_{\alpha} (\boldsymbol{r},t) =& -[H, \gamma_{\alpha} (\boldsymbol{r})](t)  \nonumber\\
=& t_{\beta\lambda}(\boldsymbol{r}'-\boldsymbol{r}'') [ \gamma_{\alpha} (\boldsymbol{r}),\gamma_{\beta}(\boldsymbol{r}')\gamma_{\lambda}(\boldsymbol{r}'')](t)  \nonumber\\
&+V_{\beta \lambda}(\boldsymbol{r}')[\gamma_{\alpha}(\boldsymbol{r}), \gamma_{\beta} (\boldsymbol{r}') \gamma_{\lambda}(\boldsymbol{r}')](t)  \nonumber \\
=& 2\gamma_{\beta}(\boldsymbol{r}')\left( t_{\alpha\beta}(\boldsymbol{r}-\boldsymbol{r}') - t_{\beta \alpha}(\boldsymbol{r}'-\boldsymbol{r})  \right)  \nonumber \\
+& 2\gamma_{\beta}(\boldsymbol{r'})\left( V_{\alpha\beta} - V_{\beta \alpha}  \right)\delta(\boldsymbol{r}-\boldsymbol{r'})
 \nonumber \\
=& \gamma_{\beta}(\boldsymbol{r}') \left(  h^t_{\alpha \beta}(\boldsymbol{r},\boldsymbol{r}')+h^V_{\alpha \beta}(\boldsymbol{r},\boldsymbol{r}') \right) \nonumber \\
=& \gamma_{\beta}(\boldsymbol{r}') h_{\alpha \beta}(\boldsymbol{r},\boldsymbol{r}').
\end{align}

Using the definition of the Green's function and the evolution of Majorana operators, we can calculate the time evolution of the retarded Green's function
\begin{align}
    i \partial_t G^r_{\alpha\beta}(t,t';\boldsymbol{r},\boldsymbol{r'}) &= \delta(t-t')\delta(\boldsymbol{r}-\boldsymbol{r'})\delta_{\alpha\beta}  \nonumber\\
    &+h_{\alpha\nu}(\boldsymbol{r},\boldsymbol{r''})G^r_{\nu\beta}(t,t';\boldsymbol{r''},\boldsymbol{r'}),
\end{align}
which can be rewritten as the Dyson equation

\begin{align}
    G^r_{\nu\beta}(t'',t';\boldsymbol{r'''},\boldsymbol{r'})=g^r_{\nu\beta}(t'',t';\boldsymbol{r'''},\boldsymbol{r'})  \nonumber\\
    +\int g^r_{\nu\alpha}(t'',t;\boldsymbol{r'''},\boldsymbol{r})h^V_{\alpha\lambda}(\boldsymbol{r,r'''})G^r_{\lambda\beta}(t,t';\boldsymbol{r''},\boldsymbol{r'})dt\,.
\end{align}
By Fourier transforming into the frequency domain we obtain
\begin{align}
G^r_{\alpha \beta}&(\omega;\boldsymbol{r},\boldsymbol{r}') =g^r_{\alpha \beta}(\omega;\boldsymbol{r},\boldsymbol{r}')  \nonumber \\
&+\sum_{\gamma\lambda;\boldsymbol{r''}} \, g^r_{\alpha \gamma}(\omega;\boldsymbol{r},\boldsymbol{r}'') h^V_{\gamma \lambda}(\boldsymbol{r''}) G^r_{\lambda \beta}(\omega;\boldsymbol{r''},\boldsymbol{r}').
\end{align}

By iteratively expanding the Dyson equation we can write the average of the retarded Green's function as 
\begin{align}
    \expval{G}= \expval{g}+\expval{gh^Vgh^VG},
\end{align}
where we have set the term linear in $h^V$ to zero since it can be absorbed into the hopping strength. Multiplying with $g^{-1}$ from the left and $G^{-1}$ from the right gives us the self energy

\begin{align}
\begin{aligned}
    \Sigma_{\alpha\beta}(\boldsymbol{r-r'}) &= \expval{h^Vgh^V}  \\
    &= \expval{h^V_{\alpha\lambda}(\boldsymbol{r,r''})h^V_{\delta\beta}(\boldsymbol{r''',r'})}g_{\lambda\delta}(\omega,\boldsymbol{r''-r'''})  \\
    &=F_{\alpha\lambda\delta\beta}g_{\lambda\delta}(\omega,\boldsymbol{r-r'}),
\end{aligned}
\end{align}
where we used that $h^V$ depends on the position as a delta function. This can be used to calculate the self-energy iteratively by using a self-consistent Born approximation:
\begin{align}
    \Sigma_{\alpha\beta}(\boldsymbol{r-r'}) = F_{\alpha\lambda\delta\beta} G^r_{\lambda\delta}(\omega,\boldsymbol{r-r'}).
\end{align}

The constants $F_{\alpha\beta\alpha\beta}$ are defined as the correlation between the disorder between different bonds
\begin{align}\label{eq:F_def}
    F_{\alpha\lambda\delta\beta}= 16\expval{V_{\alpha\lambda}(\boldsymbol{r})V_{\delta\beta}(\boldsymbol{r'})}.
\end{align}
% Note that correlations between nearest neighbor bonds are necessarily zero for time-reversal invariant systems.  
 We also assume that the correlations in Eq.~\eqref{eq:F_def} vanish for bonds in different unit cells, i.e., $\mathbf{r}\neq \mathbf{r'}$. 
From this definition, one can readily see that $F$ acquires a minus sign when flipping one of the bond directions
$F_{\alpha\beta\gamma\delta} = -F_{\beta\alpha\gamma\delta}$, and that swapping bonds is equivalent to complex conjugation 
$F^*_{\alpha\beta\gamma\delta}= F_{\gamma\delta\alpha\beta}$. 
For all bonds except adjacent ones, the disorder operators commute, and one finds $F_{\alpha\beta\gamma\delta} = F_{\gamma\delta\alpha\beta}$. 
For adjacent bonds, we get an extra minus sign from the swapping due to the Majoranas, such that $F_{\alpha \beta \beta \delta}$ must be imaginary.

Going to momentum space we write the self-energy as
\begin{align}
    \Sigma_{\alpha\beta}(\boldsymbol{k}) = F_{\alpha\lambda\delta\beta} G^r_{\lambda\delta}(\omega,\boldsymbol{k}),
\end{align}
which can be used in a self-consistent Born approximation to iteratively calculate the Green's function of the system. The spectral function, which carries information about the density of states, is related to the Green's function as 
\begin{align}
    A(\bold k,\omega) = -2\Im \Tr(G(\bold k,\omega)).
\end{align}
The imaginary part of the eigenvalues of the effective Hamiltonian $H_{eff}(k) = H_0(\bold k) + \Sigma(\bold k,0)$ broadens the single particle states of the system without disorder.

\begin{figure}[t]
	\centering
	\includegraphics[width=\columnwidth]{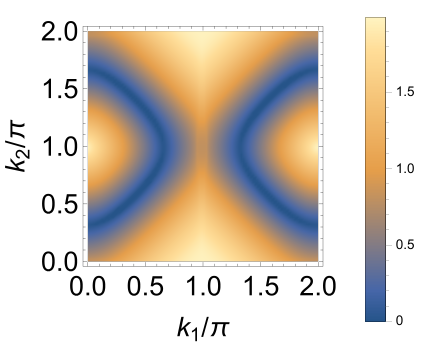}
 \caption{The real part of the energy difference between the lowest two bands.}
 \label{fig:spec1234disorderGap}
\end{figure}

To illustrate the effects of disorder  In Figs. \ref{fig:spec1234disorder}, \ref{fig:spec1234disorder2} and \ref{fig:spec1234disorderGap} we used the following values for the disorder strength
\begin{align}
    F_{1212} = -0.12, \quad F_{2323} = -0.0533, \nonumber\\
    F_{3434} = -0.03, \quad F_{1414} = -0.0833, \nonumber\\
    F_{1234} = 0.016133, \quad F_{1423} = -0.0133,\nonumber\\
    F_{1214} = -0.05 i, \quad F_{2334} = 0.04 i \nonumber\\
    F_{1423} = -0.06 i,\quad F_{1434} = 0.03 i,
\end{align}
where all the terms on the square consisting of sites 5, 6, 7, and 8 are set to zero, to break translation symmetry. 
Where the values are chosen randomly to avoid any accidental symmetries. 
% The values of $F$ is related to the disorder strengths on the bonds as $F = \frac{a^2}{3}$

\section{Anti-pseudo-Hermitian symmetry}\label{Section:Anti-ph}

\begin{figure}[h]
	\centering
	\includegraphics[width=\columnwidth]{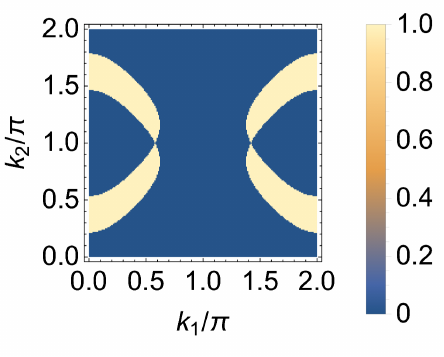}
	\caption{
 Self-symmetric states obtained from Eq.~\eqref{eq:sym_state} shows the Fermi area of the model with $\phi = 1/2 = -\theta$.}
\label{fig:fig3}
\end{figure}

Here, we derive Eq.~\eqref{eq:sym_state} that shows that chiral symmetry maps eigenstates to themselves for momenta within the Fermi area and to orthogonal states for momenta outside of the Fermi area. 
Eq.~\eqref{eq:sym_state} is used to visualize the Fermi area in Fig.~\ref{fig:fig3} and Fig.~\ref{fig:fig4} b)

Starting from the symmetry \eqref{eq:chiral_symmetry1} and acting on a right eigenstate of the non-Hermitian Hamiltonian, one finds
\begin{align}
        &U H^\dagger(\mathbf k) U^{\dagger} \ket{\psi^R_\mu} = -E_\mu \ket{\psi^R_\mu}.
        \label{eq:eigenvalue_sym}.
\end{align}
Acting with $U^\dagger$ on the left we obtain
\begin{align}
        H^\dagger(\mathbf k) U^{\dagger} \ket{\psi^R_\mu} = -E_\mu U^{\dagger}\ket{\psi^R_\mu} ,
\end{align}
or equivalently
\begin{align}
        &\left(\bra{\psi^R_\mu} U\right)H^\dagger(\mathbf k) = -E_\mu^* \left(\bra{\psi^R_\mu} U\right) .
\end{align}
The eigenvalue equation above relates, up to a constant, each right eigenstate to a left eigenstate, which we assume biorthogonalized,
\begin{align}
    U^{\dagger}\ket{\psi^R_\mu} = z \ket{\psi^L_\nu},
\end{align}
with the opposite real part of the energy, $E_\nu = -E_\mu^*$. The complex number $z$ appears because the eigenstates themselves are not normalized, and in general, $\mu \neq \nu$. However, if the eigenstate is itself symmetric,  $\mu = \nu$, its energy has to fulfill $E_\mu = -E_\mu^*$ or in other words, $\Re[E_\mu]=0$.

To confirm that the Fermi area states are protected by the symmetry, we compute the overlap 
\begin{align}
    \bra{\psi^R}U^{-1}\ket{\psi^R} \frac{\norm{\psi^L}}{\norm{U^{-1}\ket{\psi^R}}} 
\end{align}
which is quantized to $0$ for general states and $1$ for the self-symmetric states in the Fermi area. 
The normalization factor is needed to get rid of the constant $z$. 

\section{Short introduction to the Kitaev model}\label{app:KSL}
The Kitaev model introduced in Ref.~\cite{kitaev2006anyons} is one of the rare instances of an exactly solvable spin model that harbors a variety of interesting quantum spin liquid phases. 
It features spin-$\frac 1 2 $ local moments on a tri-coordinated lattice (originally the honeycomb lattice), interacting with their three nearest neighbors via a bond-directional Ising interaction:
\begin{align}\label{eq:Kham}
    \hat H &= \sum_{\langle i,j\rangle \in \gamma-\mbox{\tiny bonds}} J_\gamma \sigma_i^\gamma \sigma_j^\gamma, 
\end{align}
 i.e., depending on the bond direction, the spins aim to align their $\hat x$,$\hat y$ or $\hat z$ components, see Fig.~\ref{fig:Model1Xband} for an illustration of the square-octagon lattice. In the following, we will refer to a nearest-neighbor interaction Ising interaction with Ising axis $\hat x$ as an $x$-bond, and similarly for $\hat y$ and $\hat z$.
A very thorough and pedagogical treatment of how to solve the Kitaev model in general can be found in Kitaev's original paper \cite{kitaev2006anyons}. Here, we will review the most important steps and the relevant physics emerging on the square-octagon lattice. 

The model is exactly solvable on any tri-coordinated lattice as long as every spin interacts with its neighbors via exactly one $x$, one $y$, and one $z$ bond. This ensures that there are a macroscopic number of conserved quantities, one for each plaquette of the lattice.
We can define a plaquette operator 
\begin{align}\label{eq:Plaq_op}
    \hat W_p &=\prod_{\langle i,j\rangle_\gamma \in p} \sigma_i^\gamma \sigma_j^\gamma, 
\end{align}
where $\langle i,j\rangle_\gamma$ are the bonds in the plaquette $p$. 
For instance, for the square consisting of atoms 1 to 4 in Fig.~\ref{fig:Model1Xband}, we find the plaquette operator
\begin{align}
    \hat W &=(\sigma_1^x \sigma_2^x)(\sigma_2^y \sigma_3^y)(\sigma_3^x \sigma_4^x)(\sigma_4^y \sigma_1^y). 
\end{align}
Note that the plaquette operators commute with each other, as well as with the Hamiltonian. 
Thus, for each plaquette in the lattice, we find a conserved quantity. 
Since $\hat W_p^2 =1$ for plaquettes of even length, we can identify the corresponding eigenvalue $w_p$ with a $Z_2$ flux through the plaquette.
Determining which flux configuration allows the lowest energy is, in general, a non-trivial problem that requires extensive numerics~\cite{nasu2014finite}. 
For certain lattices, however, including the square-octagon lattice, a theorem by Lieb \cite{lieb_flux_1994} can be used to determine that the ground state sector is obtained by choosing $w_p=1$ (i.e., 0 flux) for plaquettes of length 2 mod 4, whereas plaquettes of length 0 mod 4 require $w_p=-1$ (i.e., $\pi$ flux). 
For the square-octagon lattice, Lieb's theorem can be used and the lowest flux sector has $\pi$ flux through each square and octagon \cite{Yang2007mosaic}.
The hyperoctagon lattice, whose plaquettes all have length 10, does not fulfill the requirements for Lieb's theorem. Nevertheless, the ground state resides in the zero flux sector, see, e.g., \cite{eschmann_thermodynamic_2020}.

A standard way to solve the Kitaev model is to rewrite the spins in terms of Majorana fermions
\begin{align}
    \sigma_j^\gamma &= i a_j^\gamma c_j
\end{align}
where we introduced 4 Majorana fermions: $a_j^x$, $a_j^y$, $a_j^z$, and $c_j$ per site $j$.
This enlarges the Hilbert space, and one needs to project to the physical subspace by enforcing $a_j^x a_j^y a_j^z c_j |\xi\rangle = |\xi\rangle $ for all physical states $|\xi\rangle$ and all sites $j$.
For a thorough discussion of that issue, we refer the reader to \cite{pedrocchi2011physical}. 

Introducing a $Z_2$ variable $\hat u_{j,k} = i a_j^\gamma a_k^\gamma$ for each nearest neighbor bond $\langle j,k\rangle_\gamma$ of type $\gamma$, we can then rewrite the Hamiltonian \eqref{eq:Kham} as 
\begin{align}\label{eq:KitaevMajorana}
    \hat H &= \sum_{\langle j,k\rangle \in \gamma-\mbox{\tiny bonds}} ic_j c_k \hat u_{j,k}. 
\end{align}
Note that the $\hat u_{j,k}$'s commute with each other and with the Hamiltonian.
In fact, they are directly related to the conserved quantities \eqref{eq:Plaq_op} introduced earlier.
We can thus interpret the effective Hamiltonian as Majorana fermions hopping in a \emph{static} emergent $Z_2$ gauge field (which moreover turns out to be gapped!). 
The latter is a non-interacting system and can easily be solved. 

For the square-octagon, the ground state sector has $\pi$ flux per plaquette, and the corresponding Majorana system is gapped, see Ref.~\cite{Yang2007mosaic}. 
However, by adding additional terms to the Hamiltonian that penalize $\pi$-flux, one can instead force the $0$-flux sector to be the ground state, see Ref.~\cite{Lai2011su2}. 
The latter harbors a Majorana metal, with two Majorana Fermi lines. 
Time-reversal symmetry and/or inversion symmetry ensures that the two surfaces are perfectly nested and can be translated onto each other by half a reciprocal lattice vector (see \cite{obrien2016classification} for a general discussion of such `projective symmetries'). 
Enlarging the unit cell from containing 4 to 8 sites (see Fig.~\ref{fig:Model1Xband}), we obtain a system with a single, doubly-degenerate Fermi line --- the starting point in the discussion of Sec.~\ref{section:2d_model}.

The Kitaev model on the three-dimensional hyperoctagon lattice was discussed in detail in Ref.~\cite{Hermanns2014quantum}. 
For further information about the lattice itself, we refer the reader to the original publication. 
For this manuscript, it is sufficient to know that the ground state resides in the zero flux sector and harbors two Majorana Fermi surfaces that are perfectly nested. 
The perfect nesting condition can only be destroyed by breaking time-reversal symmetry \cite{obrien2016classification}.
Similarly to the square-octagon model above, we can map the Fermi surfaces onto each other by enlarging the unit cell to contain 8 sites instead of the original 4. 
This doubly-degenerate Majorana Fermi surface is the starting point of the discussion in Sec.~\ref{sec:3Dmodel}. 

\section{Symmetries of the square-octagon model}\label{app:symmetries}
As the square-octagon model has quite a number of symmetries, which moreover are implemented differently for the 4 vs the 8-site unit cell, we will give a short overview of the relevant symmetries and their physical consequences. 

\subsection{4-site unit cell}
The square-octagon lattice is a bi-partite lattice. However, the bi-partiteness is not compatible with the 4-site unit cell, in the sense that if we choose e.g. sites 1 and 3 to be sites in the A-sublattice for one unit cell, in all the four neighboring unit cells sites 1 and 3 need to be in the B-sublattice. 
The Kitaev model on a bi-partite lattice is invariant under sublattice symmetry, which is implemented by 
\begin{align}
c_j(\mathbf{r})&\rightarrow\left\{ \begin{array}{cc}
     c_j (\mathbf{r}) & \mbox{ for $\mathbf{r} \in $ sublattice A}  \\
     - c_j (\mathbf{r})& \mbox{ for $\mathbf{r} \in $ sublattice B} 
\end{array}\right. .
\end{align}
Since the bi-partiteness is not compatible with the 4-site unit cell of the square-octagon lattice, the corresponding transformation becomes non-local in momentum space: 
\begin{align}
  \hat h(\mathbf k) &= -U_{SLS} \hat h(\mathbf k + \mathbf k_0) U_{SLS}^{-1}
\end{align}
where $\mathbf k_0$ is half a reciprocal lattice vector and $U_{SLS}$ is a unitary matrix.
This feature has profound consequences for the physical properties of the Kitaev spin liquid as explained in Ref.~\cite{obrien2016classification}. 
In particular,  implementing time-reversal symmetry for Majorana systems requires a sublattice transformation in the gauge sector. 
Thus, time-reversal inherits the same transformation symmetries from sublattice symmetry: 
\begin{align}
    \hat h(\mathbf k) &= U_{TR} \hat h^*(-\mathbf k + \mathbf k_0) U_{TR}^{-1}, 
\end{align}
where $U_{TR}$ is a  unitary matrices.  
The non-trivial transformation under time-reversal immediately implies that the stable zero modes for this model form lines, i.e. the Majorana Fermi lines seen in Fig. \ref{fig:fig1}. 

Apart from several mirror symmetries, and a four-fold rotation symmetry --- both of which are not relevant for our discussion --- the square-octagon model also possesses inversion symmetry. 
The inversion center can sit in the middle of the squares, the middle of the octagons, or the middle of the z-bonds connecting two squares. 
For the 4-site unit cell, all these choices lead to the same constraint on the Hamiltonian, namely 
\begin{align}
    \hat h(\mathbf k) &= U_{I} \hat h(-\mathbf k + \mathbf k_0) U_{I}^{-1}, 
\end{align}
where $U_I$ is a unitary matrix. 
Both time-reversal and inversion symmetry imply that if there is a zero-energy state at $\mathbf k$, there must be another zero-energy state at $-\mathbf k + \mathbf k_0$, i.e. they map one of the Majorana Fermi surfaces to the other. 
Only by breaking both these symmetries can we destroy the perfect nesting condition. 

\subsection{8-site unit cell}
We now assume that translation symmetry is partially broken, such that the 8-site unit cell is appropriate. 
This unit cell is compatible with the bi-partiteness of the lattice, so both time-reversal and inversion will only map $\mathbf k\rightarrow -\mathbf k$, without an additional translation in $\mathbf k$-space. 
For the 8-site unit cell, the inversion symmetry with its center on the z-bond is automatically broken, so we only consider those with centers in the square/octagon.  

In particular, time-reversal is now given by 
\begin{align}\label{eq:TR8}
    \hat h(\mathbf k) &= U_{TR} \hat h^*(-\mathbf k) U_{TR}^{-1}
\end{align}
and inversion symmetry by 
\begin{align}
    \hat h(\mathbf k) &= U_{I} \hat h(-\mathbf k ) U_{I}^{-1}, 
\end{align}
where $U_{TR}$ and $U_{I}$ are now 8x8 unitary matrices. 
The general symmetry analysis in Ref.~\cite{obrien2016classification} implies that this implementation of time-reversal has stable zero modes at discrete points in $\mathbf k$-space, while lines of zero modes are generically unstable. 
To understand this, notice that \eqref{eq:TR8} implies that the time-reversal invariant Hamiltonian can  be brought into the form 
\begin{align}\label{eq:H_offdiag}
    \hat h(\mathbf k) &= \left(\begin{array}{cc}
        \mathbf{0} & q(\mathbf k) \\
         q^\dagger(\mathbf k) &  \mathbf{0}
    \end{array}\right). 
\end{align}
The zero modes of $\hat h$ are obtained by requiring $det(\mathbf q(\mathbf k))=0$. 
The latter are in fact two constraints, $Re[det(\mathbf q)]=0$ and $Im[det(\mathbf q)]=0$, both of which form closed lines in $\mathbf k$-space. 
The zeroes of $\hat h$ are located at the intersections of these lines.  
The existence, though not the position, of these intersections are stable against all perturbations that do not break time-reversal. 

This is the general behavior, but the existence of inversion symmetry adds a twist to the story. 
Inversion symmetry forces $det(\mathbf q)$ to be real. 
Thus,  one of the constraints mentioned above is trivially satisfied, implying that the stable zero modes instead form lines. 
One should note here, that the zero mode lines for the 4-site unit cell arise from Majorana Fermi lines crossing zero energy. 
Here, in contrast, the lines arise from two bands touching at zero energy. 
The former situation is stable to any symmetry breaking, while the latter needs inversion symmetry. 

We also want to note another twist, this time arising from the non-hermitian perturbation of the model. In the Hermitian system, breaking time-reversal necessarily implies broken sublattice symmetry. 
This is no longer the case for the non-Hermitian model. 
For instance, non-zero phases in Eq.~\eqref{eq:hamiltonian} break time-reversal, but not sublattice symmetry. 
Thus we can have the novel situation where time-reversal is broken, but sublattice symmetry is intact. 
The off-diagonal structure of the Hamiltonian in \eqref{eq:H_offdiag} relies in fact on sublattice symmetry. 
For the non-Hermitian model of Eq.~\eqref{eq:hamiltonian}, above analysis about the stable zero modes remains valid even in the absence of time-reversal. 
The effective Hamiltonian of the disordered system, on the other hand, does break sublattice symmetry explicitly, though not time-reversal. 

\bibliography{DRAFT}

\begin{thebibliography}{47}
\bibitem [{foo()}]{footnote1}%
  \BibitemOpen
  \href@noop {} {}\bibinfo {note} {{For the sake of this subsection, we simply assume that the Hamiltonian \eqref{eq:hamiltonian} was obtained from a fermionic hopping model, instead of a Majorana hopping model}}\BibitemShut {NoStop}%
\end{thebibliography}
\end{document}